# Temperature Dependence of Reconfigurable Bandstop Filters Using Vanadium Dioxide Switches

Andrei A. Muller, Matteo Cavalieri, Adrian M. Ionescu

(NanoLab), École Polytechnique Fedérale de Lausanne (EPFL), 1015 Lausanne,

Switzerland

**ABSTRACT**

In this letter we report and investigate the temperature dependency of various radio frequency parameters (RF) for a fabricated reconfigurable bandstop filter with vanadium dioxide ($VO_2$) switches measured up to 55 GHz. Here the insulator to metal (ITM) and metal to insulator transition (MIT) hysteresis of the $VO_2$ thin film influence on the RF characteristics of the filters is analyzed from 25 °C and 120 °C in heating and cooling. The resonance frequency and maximum insertion loss (IL) stability and sensitivity with temperature variations are explored. It is noticed that increasing the temperature with 50 °C from 25 °C (or decreasing it with 50 °C from 120 °C) will result in a less than 1% fractional frequency shift in respect to the off and on resonance frequencies. The sharp DC conductivity levels variations of the $VO_2$ thin film around the transition temperatures translate into sharp effects on the resonance characteristics of the filters. On the contrary, the maximum IL levels are less sensitive to the DC films sharp conductivity changes around the $VO_2$ transition temperature. Last, we see that the RF parameters in heating and cooling at 80 °C, above (but close to) the DC transition temperatures of $VO_2$ exhibit completely different resonance frequencies. The RF results reported close to the transition temperatures for the $VO_2$ thin films can diverge in heating and cooling, thus of a more insightful understanding of $VO_2$ reconfigurable RF devices has to include temperature dependent measurements at various temperatures below MIT and ITM in the RF ranges too.

The phase-change material, vanadium dioxide (VO$_2$,) exhibits a monoclinic crystal structure and behaves like an insulator below its insulating-to-metal transition temperature ($T_{ITM}$), which is around 68 °C in bulk VO$_2$[1-6]. Above this temperature the monoclinic crystal structure changes to tetragonal crystal structure. Conversely, a reversible Metal-to Insulator transition occurs at ($T_{MIT}$) a bit below 68 °C for bulk VO$_2$ when temperature is decreased. Furthermore, dimensional scaling applies, and the nature of the phase transition is additionally conserved in VO$_2$ thin films and ultra-thin films[6-7]. Unlike in single crystalline forms of VO$_2$ where both phase transitions occur around 68 °C and exhibit narrow hysteresis, in thin films the critical temperature (defined as the average between $T_{ITM}$ and $T_{MIT}$ i.e $T_c= (T_{ITM} + T_{MIT})/2$, the width of the hysteresis loop, and the amplitude of the transition depend strongly on the film morphology, doping and deposition substrate[7-8]. Due to the ultrafast transition, the switching between the insulating and metallic phases can occur very quickly, (of the order of a few phonon oscillations) and its relative closeness to room temperature, vanadium dioxide has raised tremendous interest in both fundamental research and a variety of applications in electronics. VO$_2$ has been used a switching element in the infra-red frequency bands[9-11], THz regions[12-15] while recently too in the RF frequency bands[5,16]. VO$_2$ RF switches[16] can operate under a lower transition temperature with reduced power consumption, compared with other thermally triggered switches, particularly unlike most of the other RF switches based on diodes or micro-electro-mechanical systems MEMS, the device structure and fabrication of VO$_2$ switches can be much simplified. VO$_2$ switches small size and high linearity increased its attractiveness for the RF community too. A variety of reconfigurable RF devices using VO$_2$ switches such as antennas[17], inductors[18], filters[19-22] have been recently reported. Their performance however is very much dependent on the VO$_2$ thin film depositions and conductivity levels in the insulating (off state) and conductive (on state). The thin films conductivity levels reach hundreds of thousands of Siemens (S) above $T_{IMT}$ when deposited on sapphire, while mostly just tenths of thousands of S when deposited on cheaper CMOS Si substrates[18-22], hindering the on state behavior of the devices. In the off state on the other hand, the dielectric behavior and non-zero DC conductivity levels of the thin films are also affecting the RF performances.

Here, we report a reconfigurable bandstop filter with $VO_2$ switches on a CMOS compatible Si substrate exhibiting a wider tuning range in respect to all K, Ka or V band experimentally bandstop filters reported employing $VO_2$ switches[19-22] (to the best of our knowledge). The tuning range of the filter defined as *t(%)= |fmax-fmin|/fmax* where fmax and fmin represent the resonance frequencies of the filter at 25°C and 120 °C respectively is here of 34% improving the tuning ranges reported by in [19,21,22] without deteriorating the attenuation levels. Then, unlike most studies which analyze the RF components behavior only at two temperatures[17-22], here we perform a series of measurements at a variety of temperatures on both DC films characteristics and on the RF parameters of the filters. The stability of the RF performances with temperature is extracted and the correspondence between the DC thin film temperature sensitivity and RF performance of the filters is analyzed.

The filters were fabricated using the same methodology as in standard microelectronic processes starting with a high-resistivity (10000 Ω·cm) 525 μm thick silicon substrate. A 300 nm thick amorphous silicon layer was firstly deposited, then the substrate was passivated with 500 nm $SiO_2$ deposited by sputtering. A 140 nm thin $VO_2$ film and was deposited at 400 °C in oxygen atmosphere by a Pulsed Laser Deposition (PLD) system using a $V_2O_5$ target and then annealed at 475 °C in the same system for 10 min. The film was then patterned using standard photolithography followed by dry etching. A Cr (20 nm)/Al (400 nm) bi-layer was deposited to contact the patterned VO$_2$ film. This thin contact layer allowed for the realization of 1.3 µm gaps between the contact pads. Additionally, a 2.4µm-thick Al layer was deposited on top of these contact metal by conventional lift-off methods to create the final coplanar waveguide elements. Fig. 1 (a) shows the filter schematic while Fig.1 (b) presents a cross sectional schematic of the switch. The angle θ represents the angle of an additional gap in a classical split ring topology[22], in respect to the common outer ring gap, here θ being at 5 o'clock unlike 3'o clock in[22] (considering 12 a clock where the classical outer gap lies). Fig. 2 (a) presents the switch photo and Fig 2 (b) the grain size of the VO$_2$ thin film used.

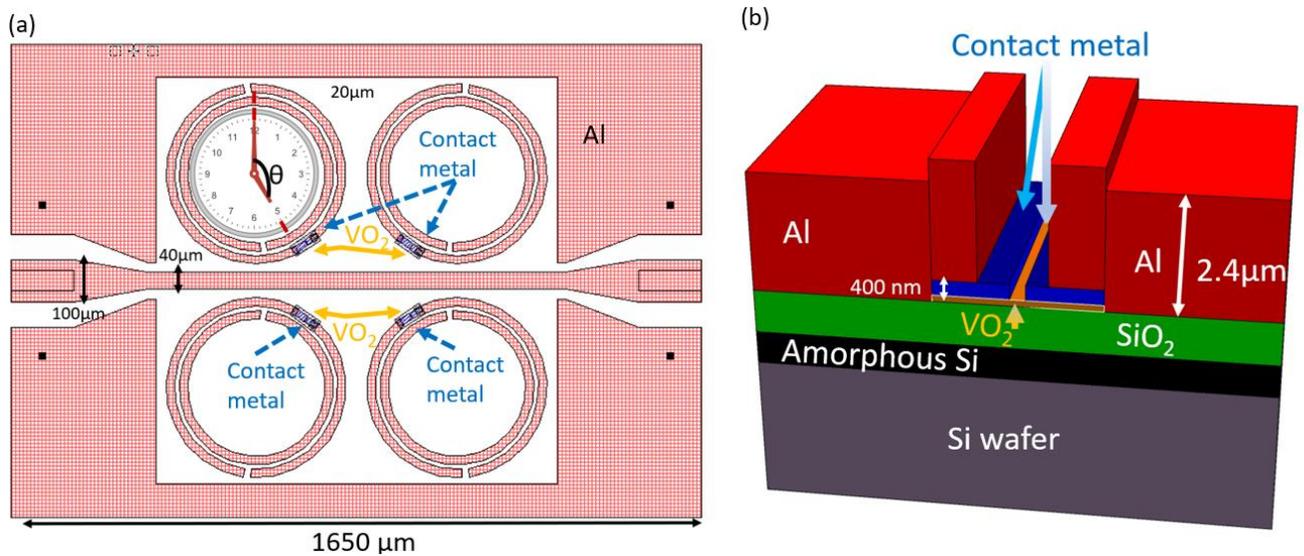

FIG. 1. (a) Schematic of the filter (b) Cross-sectional schematic of the switching elements.

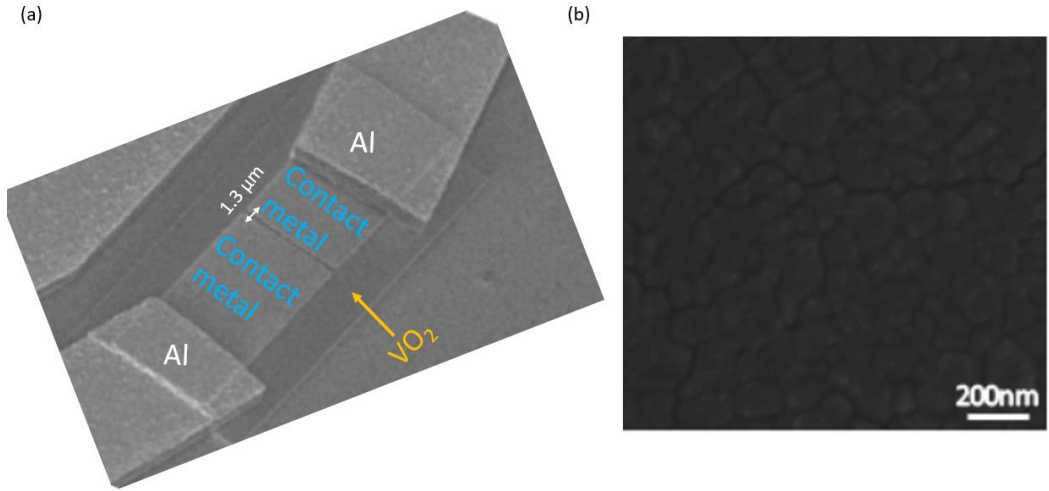

FIG. 2. (a) SEM photo of the switch (b) Grain sizes of the $VO_2$ thin film.

The filter RF performances are initially tested only at two different values (here at room 25°C temperature and then at 120°C), as in most RF devices including $VO_2$ switches[5,17-22]. The tests are performed using an Anritsu vector network analyzer (VNA) with a heating stage below the sample in order to control its temperature. Figure 3 (a) shows the filter performances in terms of $S_{21}$ (dB)- the filter exhibiting a resonance frequency of 45.64 GHz in off state ( here at room temperature) and at 29.93 in the on state (here at 120 °C). The maximum insertion loss (IL) at the off/on resonances are of 18.29 dB and 14.02 dB. The results improve in terms of reconfigurability the previous results including $VO_2$ switches for similar frequency bands exhibiting here a tuning range $t = |f_{max}-f_{min}|/f_{max} = 34$ %. A comparison in respect to previous works is displayed in Table 1, presenting the improvement in respect to our previous works[21-22] in respect to the tuning range.

Table 1 RF filter performances using $VO_2$ switches at room temperature and at a temperature above $T_{ITM}$ in heating

| References | [19] | [21] | [22] | this work |
|---|---|---|---|---|
| Tuning range (%) | 12% | 19% | 23% | 34% |
| Maximum IL (dB) | 16 and 18 | 12.8 and 18 | 14.10 and 18 | 14.02 and 18.29 |

| Resonance frequencies (GHz) | 19.8 and 22.5 | 28 and 34 | 29.74 and 38.70 | 29.93 and 45.64 |

Unlike most RF devices reported including VO₂ switches whose performances are only validated in on/ off cases[5, 17-22], here we go further with the tests in order to understand how the heat influence on the VO₂ thin film alters the overall RF performances of the devices. First, the VO₂ thin film DC conductivity is assessed using four probe measurements and the results are presented in Fig. 3 (b). The on state conductivity levels are just below 50,000 S/m while the off state values are around 30 S/m- the results are in the same trend as other depositions on SiO₂/Si wafers[21,22], the conductivity of the VO₂ being considerably affected by substrate lattice mismatches, grain size and deposition parameters[7,8,22]. Then, we perform VNA measurements at 25 °C, 50 °C, 60 °C, 66 °C, 70 °C, 75 °C, 80 °C, 83 °C, 86 °C, 90 °C, 100 °C, 120 °C in heating and cooling and get the results presented in Fig. 3(c) and Fig. 3(d).

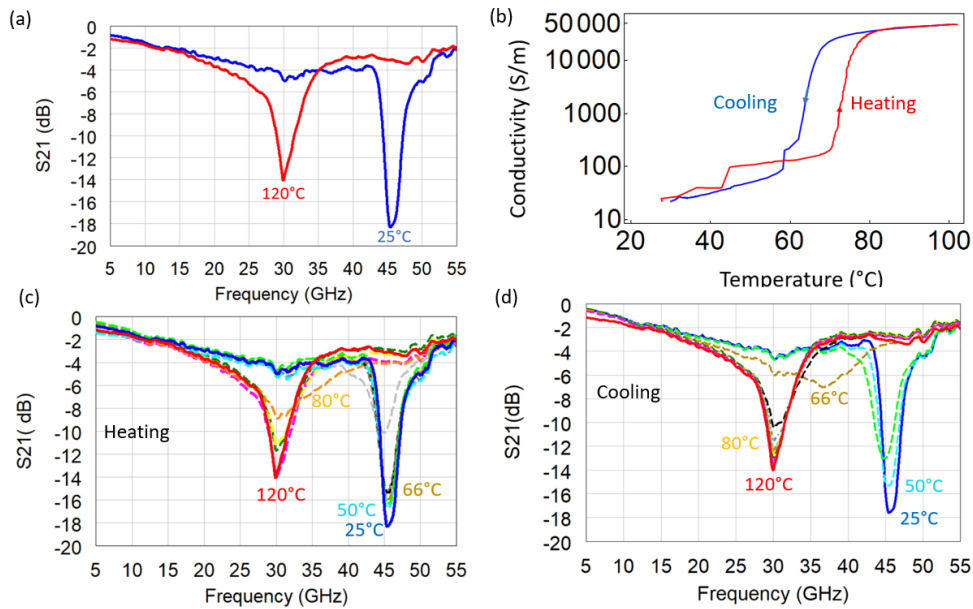

FIG. 3 (a) Filter on/off S21 (dB) parameter, (b) VO₂ thin film conductivity levels in heating and cooling. (c) S21(dB) parameter of the filter at 25 °C, 50 °C, 60 °C, 66 °C, 70 °C, 75 °C, 80 °C, 83 °C, 86 °C, 90 °C, 100 °C, 120 °C in heating (d) S21(dB) parameter in cooling.

From an RF point of view, one of the important features when it comes to a resonator is its temperature stability in respect to its resonance frequency and maximum insertion loss (IL) at the resonance frequency. In order to grasp this influence Fig 3 (c) and (d) results are arithmetically manipulated in Fig. 4 and Fig. 5 for understanding the resonance frequency ($f_r$) evolution with temperature change first and then the maximum IL at resonance.

Fig 4 (a) shows the variation of the resonance frequency once temperature is increased in respect to the off state and when temperature is decreased in respect to the on state. The graph illustrates little changes in heating up to 80 °C (and thus for a heating variation of 55 °C) and also little changes in temperature decrease up to 70 °C. In Fig. 4 (b) we express this variation in terms of fractional frequency shift in respect to the off/on resonance frequencies $\Delta f/f_{off}$ in heating and $\Delta f/f_{on}$ in cooling, where $\Delta f = f_r - f_{off}$ in heating and $\Delta f = f_r - f_{on}$ in cooling.

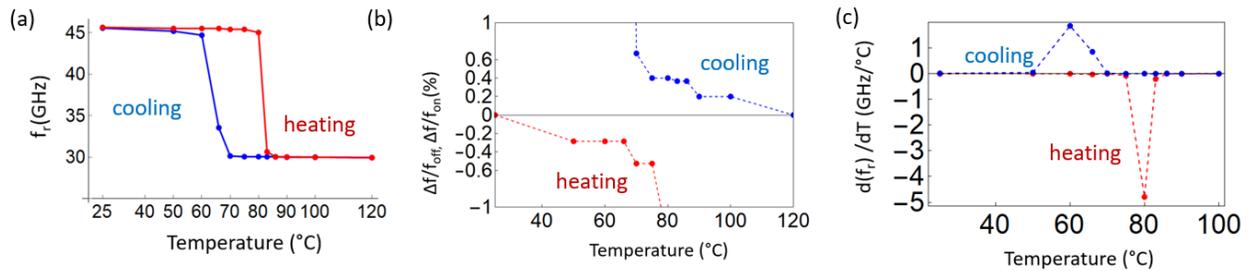

FIG. 4 In heating and cooling (a) Resonance frequency variation with temperature (b) fractional frequency shift in respect to the on and off resonance frequencies, (c) Derivatives of the resonance frequency variation in respect to temperature variation.

We may see a fractional frequency shift below 0.6 % in heating from room temperature up to 75 °C and cooling from 120 °C down to 70 °C, overall viewing that in order to keep the frequency stable with 0.6% one may resist to an increase of temperature to up to 75 °C, and while in cooling a decrease up to 70 °C. Fig. 4 (c) shows the speed of the resonance frequency change in heating and cooling presenting a relative sharp transition around $T_{ITM}$ and $T_{IMT}$ of the VO$_2$ thin film.

Fig. 5 on the hand exploits the attenuation changes with temperature increase and decrease expressed here in terms of maximum insertion loss (max IL). Fig. 5 (a) displays clearly how the maximum attenuation levels are deeply affected by the temperature changes. Fig 5 (b) illustrates the evolution of the fractional maximum insertion loss at resonance frequency variation in respect to the maximum levels in on and off states (where $\Delta IL = IL(f_r) - IL(f_{off})$ in heating and $\Delta IL = IL(f_r) - IL(f_{on})$ in cooling). One may notice a change of 17 % at an increase of temperature from 25°C to 75°C. The same is observed in cooling when the temperature is decreased to 75°C. Fig 5 (c) displays the resonance frequency change in heating and cooling showing a relative far less sharp transition around $T_{ITM}$ and $T_{IMT}$ of the VO$_2$ thin film than in the case of the resonance frequency.

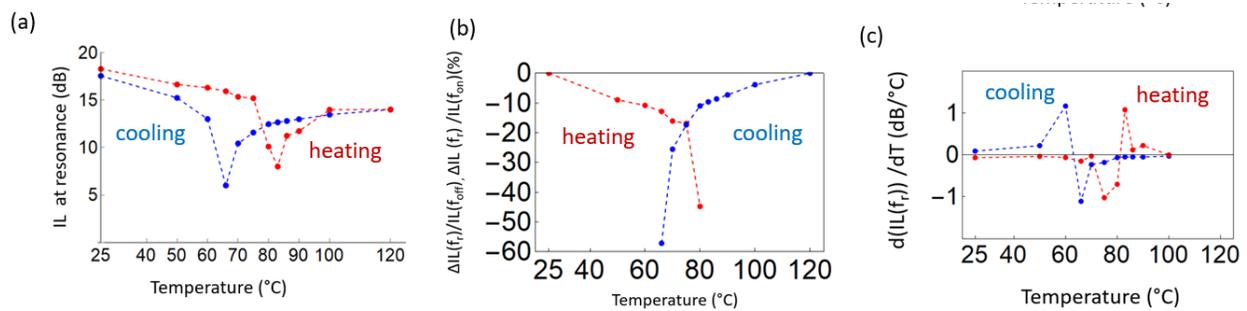

FIG. 5 In heating and cooling (a) IL variation with temperature (b) fractional IL variation in respect to the on and off resonance frequencies IL, (c) Derivatives of the IL at resonance frequencies variation in respect to temperature variation.

Finally, Fig. 6 (a) displays the conductivity increase and decrease rate with temperature of the VO$_2$ DC film, while Fig. 6 (b) shows the fractional variation of resonance frequency and attenuation with respect to the off state performances, below $T_{ITM}$. Fig. 6 (c) indicates the fractional variation of the resonance frequency and attenuation from on state, decreasing the temperature down to the $T_{MIT}$.

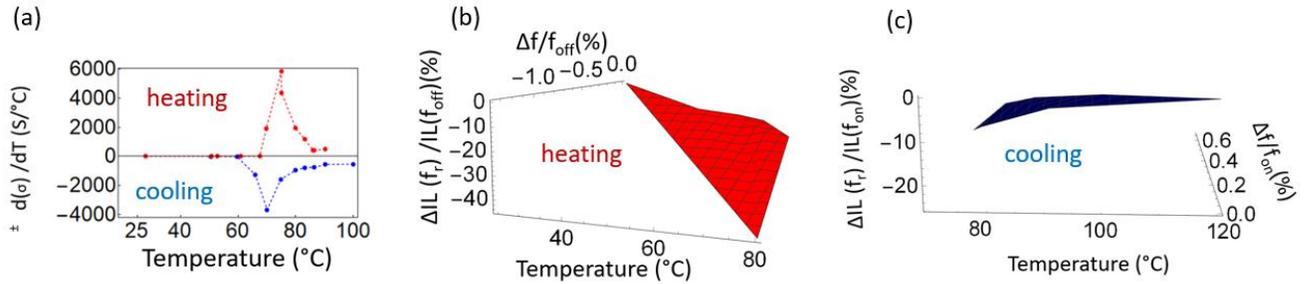

FIG. 6 In heating and cooling (a) IL variation with temperature (b) fractional IL and resonance frequency variation in respect to the off state ones below $T_{ITM}$, (c) fractional IL and resonance frequency variation in respect to the on state ones above $T_{MIT}$.

Last, we would like to point out that the measured RF parameters, may have completely different responses in heating and cooling also when the temperature is above the $T_{ITM}$ of the VO$_2$ thin film. Authors (as us too[18, 21-22]) tend to present the results of the reconfigurable RF components in the off state around room temperature and in on state, usually at a single temperature above $T_{ITM}$[5, 12, 17-23] in heating. Seldom, the behavior of the RF devices at other temperatures is checked, with few exceptions in discrete frequency points[16] in heating and cooling, while on a broadband frequency range but only in heating and up to 80 °C[24], Table 2 showing briefly the RF devices testing tendency in literature. Fig. 7 illustrates the uncertainty of the measured filter S21 parameters in cooling and heating at 80°C above $T_{ITM}$ of the thin film which is around 71 °C.

Table 2 RF VO$_2$ based reconfigurable components testing temperatures in heating and or cooling

|  | Temperatures | heating | cooling | comment |
|---|---|---|---|---|
| 5 | 27°C/ 68°C | yes | - | Simulation, 2 temperatures, off/on |

| 12 | below 68°C/above 68°C | yes | - | 2 temperatures, off/on |
|---|---|---|---|---|
| 15 | below 68°C/above 68°C | yes | - | 2 temperatures, off/on, we refer to the results presented in the article overlapping the RF band in between 100 GHz-300 GHz |
| 16 | 30°-90 °C | yes | yes | in discrete frequency points the RF switches are tested at more temperatures |
| 17 | 21°C/90 °C | yes | yes | 2 temperatures, off/on |
| 18,21,22 | 25°C/100 °C | yes | - | 2 temperatures, off/on, us in previous works |
| 19 | Room temperature/80 °C | yes | - | 2 temperatures, off/on |
| 23 | Room temperature/85 °C | yes |  | 2 temperatures, off/on |
| 24 | 20°C-80 °C | yes | - | Measured in heating |

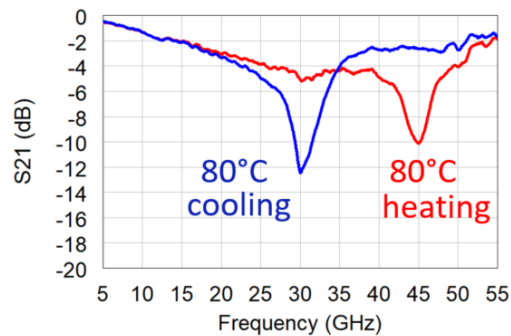

FIG. 7 In heating and cooling the filter: measurement for the S21 parameters above the $T_{ITM}$ of the VO$_2$ thin film at 80°C- dB plot.

In conclusion we have studied and discussed the RF performances temperature dependency for a reconfigurable bandstop filter using VO$_2$ switches . Further, we made to the best of our knowledge for the first time a thermal sensitivity and stability analysis of the RF filters with VO$_2$ switches below the transition temperature and above. It was seen that while the resonance frequency of the filters follow the sharp transition of the $VO_2$ DC film conductivity characteristics, not the same is the case in the terms of bandstop rejection, which is very much affected by each conductivity level change. Moreover, we have noticed that in order to keep the fractional resonance frequency variation below 1 % one can increase in the off state the temperature to up to 75°C, then a sharp shift of 34% is observed. Similar results were observed for the off state, these results conclude that although temperature stability is very important for keeping high bandstop rejection levels, not the same is the case for the resonance frequencies. Last we have shown that in the case of RF filters with $VO_2$ switches, their behavior may be completely different even above the $T_{ITM}$ of the $VO_2$. We attested too that in order to predict the RF performance of the filters with $VO_2$ switches one needs or to know the history of the RF device, or else to maintain the temperature considerably above their $T_{ITM}$, or considerably below their $T_{MIT}$ otherwise divergent results can occur in the RF measurement in heating and cooling.

**ACKNOWLEDGEMENTS**

This work was supported by the HORIZON 2020 FET OPEN PHASE-CHANGE SWITCH Project under Grant 737109. Andrei Muller would also like to ack. Riyaz A. Khadar for the help in the fabrication flow.

[Insert Running title of <72 characters]